# Hyper Heterogeneous Cloud-based IMS Software Architecture: A Proof-of-Concept and Empirical Analysis

Pascal Potvin[1,2], Hanen Garcia Gamardo[1,2], Kim-Khoa Nguyen[2,] Mohamed Cheriet[2]

[1]Ericsson Canada inc., Town of Mount-Royal, Canada
{pascal.potvin, hanen.garciagamardo}@ericsson.com
[2]École de Techologie Supérieure, Montréal, Canada
{knguyen}@synchromedia.ca
{mohamed.cheriet}@etsmtl.ca

**Abstract.** The IP Multimedia Subsystem (IMS) defined by the 3GPP has been mainly developed and deployed by telephony vendors on vendor-specific hardware. Recent advances in Network Function Virtualisation (NFV) technology paved the way for virtualized hardware and telephony function elasticity. As such, Telecom vendors have started to embrace the cloud as a deployment platform, usually selecting a privileged virtualization platform. Operators would like to deploy telecom functionality on their already existing IT cloud platforms. Achieving such flexibility would require the telecom vendors to adopt a software architecture allowing deployment on many cloud platforms or even heterogeneous cloud platforms. We propose a distributed software architecture enabling the deployment of a single software version on multiple cloud platforms thus allowing for a solution-based deployment. We also present a prototype we developed to study the characteristics of this architecture.

## 1 Introduction

The IMS [1] is a standardized solution that addresses an operator's need to provide advanced services on top of both mobile and fixed networks. It uses the Session Initiation Protocol (SIP) to establish and manage sessions. Fig. 1.A presents a view of the IMS as it is currently standardized. We consider the simplified view of the IMS with its main functions; Call Session Control Functions (CSCF), Home Subscriber Server (HSS), Multimedia Telephony (MMTEL) and Media Resource Functions (MRF) circled in Fig 1.A. Current IMS deployments are typically done on vendor-specific hardware. For example, Ericsson has a family of hardware platforms [2] for IMS deployment purposes. In other words, IMS functions are customarily deployed on dedicated physical nodes. Fig. 1.B shows a possible deployment of the core IMS functionality on server racks.

The Network Function Virtualisation (NFV) standardization effort [3] has recently sought to introduce virtualization platforms for telephony functions and IMS. The

NFV standard leverages the evolution of the current (predominantly) vendor-based hardware deployment consisting of Physical Network Functions (PNF) to a vendor-agnostic hardware platform running on virtualized hardware with Virtual Network Functions (VNF). NFV introduces the concept of elasticity for telephony application deployment, allowing a wide range of potential implementations of the elasticity concept from no implementation at all to fully automated.

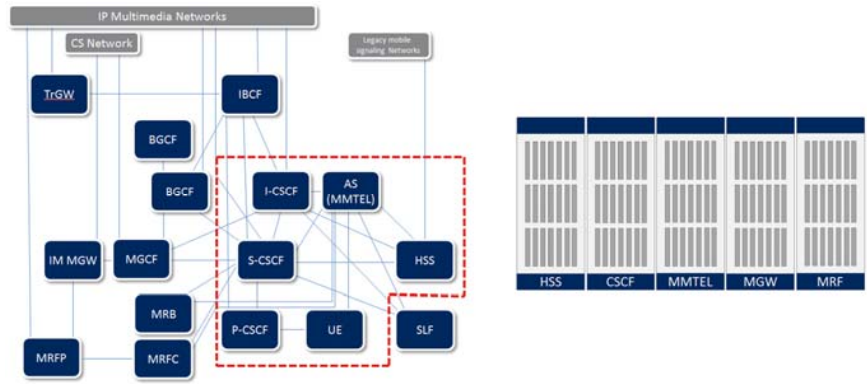

**Fig. 1.** (A) The IP Multimedia Subsystem (IMS) with the simplified view we consider being circled; (B) A possible IMS deployment on server racks.

Until very recently, the deployment of a VNF was still executed on a per node basis, thus providing coarse scalability and limited elasticity [4]. The problems associated with such coarse scalability are well covered in [5] and the general problem of scaling the IMS [6] is considered in [4] and [7]. Prior solutions focused on resource over-provisioning to solve scalability issues leading to poor resource utilization derived from scaling on a per-node basis. A dynamic distribution, or concentration of IMS functionality has been proposed [8], but this still maintains node-based coarse scaling. This approach helps increase utilization but fails to solve the over-provisioning issue. A similar approach, so called "Merge-IMS" [9] proposes a pool of IMS VMs containing CSCF and HSS functionality whereby a specific VM instance is assigned to a subscriber at registration.

Today, Cloud providers usually build their cloud on homogeneous commoditized hardware to reduce acquisition and operating costs. As Cloud technology is being adopted, Telecom vendors might have to deploy their software on an operator's cloud which is very different from one operator to another. Part of the Telecom vendor's software functionality might be better suited for certain types of hardware. Given this, the ability to deploy the same software in a solution-defined heterogeneous pool of computing resources is desirable.

The remaining sections of this paper are organized as follows. Section 2 presents previous work related to our research. Section 3 describes the three main layers of the "Hyper Heterogeneous Cloud" architecture we named Unity: i) the Unity architecture and the Unity framework, ii) the re-designed IMS application running on the Unity framework, and iii) the hardware platform used for our implementation and experimentations. Section 4 describes our experiment and finally in Section 5 we discuss our findings and conclusions.

## 2   Related Work

So far, the definition of "Heterogeneous Cloud" is still unclear. Some authors associate it to the Cloud software stack currently being built by multiple vendors [10] e.g. a management tool from one vendor driving a hypervisor from another. Others associate it to the use of hardware clusters that contain heterogeneous equipment [11],[12] e.g. general purpose computing platforms sitting next to specialized accelerators or mixed-characteristic general computing platforms where some equipment has faster processing, better  I/O capacity or provides different memory/storage capacities. Nevertheless, much work has been done on the Heterogeneous Cloud. In [11] the authors propose a solution to schedule tasks to best fit hardware computing resources; in [12] the authors propose a cloud built of a mix of Central Processing Unit (CPU) and Graphical Processing Unit (GPU) based computing resources through virtualization. Another CPU/GPU study [13] looks at how proper allocation and scheduling on such heterogeneous cloud can benefit Hadoop [14] workload.

Unfortunately, to the best of our knowledge, no architecture has yet been proposed for the telecom sector in order to provide portability between multiple cloud environments or to enable solution-oriented heterogeneous cloud deployments. At the same time, no approach has been proposed to distribute and instantiate core IMS functionality in an on-demand, per subscriber and per service basis. This paper is therefore dedicated to address the following questions: i) Can we define a cloud-based software architecture that can be easily deployed on heterogeneous hardware clusters (containers, virtual machines, bare metal servers clusters, specialized accelerator clusters…)?, ii) Can we implement an IMS over such an architecture in order to provide on-demand per subscriber and per service functionality?, and finally, iii) what would be the characteristics of such an architecture and how does it compare to a node-based deployment?

## 3   Unity Cloud

The "Hyper Heterogeneous Cloud" architecture named Unity that we propose in this paper can be deployed on a set of different hardware infrastructures, using a mix of management tools and a mix of deployment technologies. In other words, part of the deployment may be on Virtual Machines (VMs), on containers and on bare metal to take advantages of the various platforms and their availability. Our goal is to build a system and software architecture which allows a single software base to be deployed on heterogeneous hardware and cloud platforms. Specific requirements are met through deployment configuration rather than a design for a specific platform set.

To study the characteristics of a Hyper Heterogeneous Cloud Deployment Software Architecture, we built a simplified IMS system on a Microservices-based architecture [15]. This gives us the flexibility to distribute the IMS functions on a combination of platforms, through a Descriptor file which defines the available pools of platform resources and the deployment model of the Microservices. The Meta Manager and Orchestrator we built can deploy the functionality on heterogeneous platforms. This approach allows defining Hybrid deployments since the defined platforms could as well be provided by a Public Cloud. The list of Microservices developed for the Unity Cloud and the IMS functionality implemented is detailed later in this paper. We first focus on the software architecture and infrastructure enabling a Hyper Heterogeneous Cloud deployment.

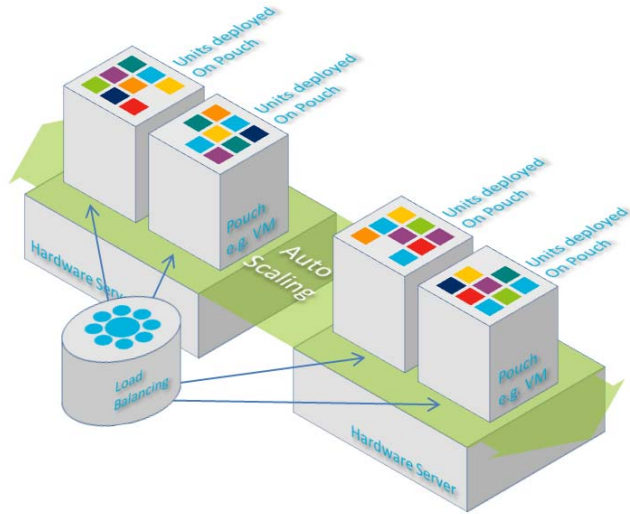

**Fig. 2.** Distributed deployment of Units on Pouch scaling as needed.

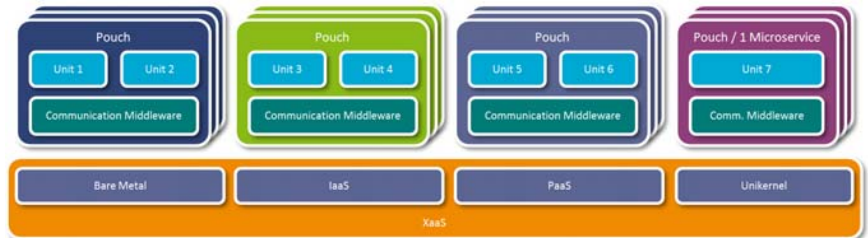

**Fig. 3.** Concept of Pouch deployed on XaaS.

In this architecture the Cloud platform is responsible for allocating computing, network and storage resources to provide the required telecom functionality on a per-user or per-service basis. In order to cater to the heterogeneity of the platforms (PaaS, IaaS, BareMetal, etc.) we introduce an abstraction layer which represents an instance of a computing resource on a platform. We define a Pouch (Fig. 2 and Fig. 3) as a computing resource combined with a lightweight platform framework. The framework supports functions which are offered as a library to the application code rather than an over the network as a service. The Pouch can be seen as a set of libraries and daemons running on a computing resource to support the Microservices and facilitate access to other services. In practice a Pouch can be a Bare Metal server, a Virtual Machine on IaaS, a Container/job on PaaS, a Microservice on a Unikernel, etc. The number of Microservices and instances held by a Pouch can vary from one to thousands depending on the characteristics of the host where the Pouch is deployed.

One can scale out any number of Pouches on a platform; a Unit which can be assimilated to an actor in the Actor Model and instantiated within a Pouch is able to transparently communicate with other Units within other Pouches through the Communication Middleware.

### 3.1 The Unity Cloud Architecture

The Unity architecture (Fig. 4) defines a set of functionality or services allowing a Microservices-based application to be deployed on Hyper Heterogeneous Cloud platforms. The Microservices performs a specific task and covers a single scaling domain. For example a Microservice may handle a limited number of related telephony services or the HSS interrogating functionality of an application. The Microservices are deployed as "Units" as follows.

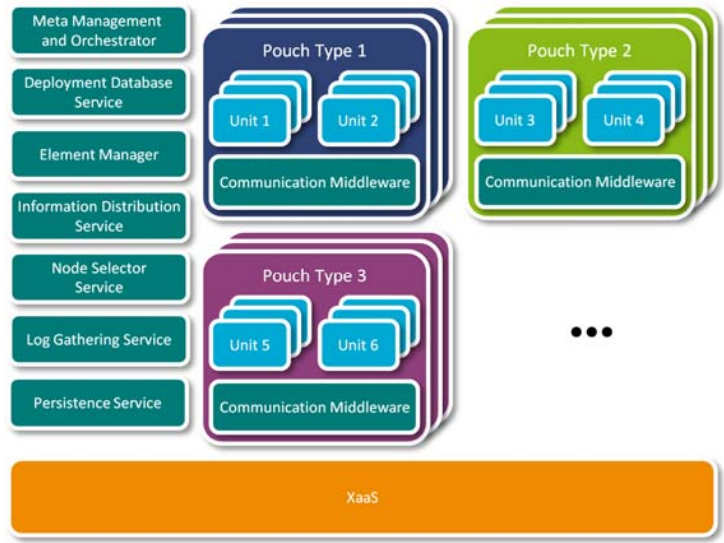

**Fig. 4.** Unity Cloud Architecture.

**Meta Management and Orchestrator (MMO)** is responsible for reading the Descriptor file and deploying the appropriate Pouches on the available platform pools. It also monitors usage information from the CMWs via the IDS and instantiates new Pouches to provide elasticity.

**Element Manager (EM)** is responsible for configuring the Microservices (called Units) running on the Pouches.

**Communication Middleware (CMW)** forms the basis of the Unity Cloud Architecture; each Pouch is required to run a single instance of a CMW. It manages most basic functionality that is required for the Unity Cloud operation such as Inter-Unit communication, Unit spawning, Pouch monitoring and Unit/Service address resolving.

**Node Selector Service (NSS)** implements the logic of spreading the Subscribers' service instances on the available Pouches. It ensures that most of the service requests for a Subscriber are made to the same Pouch in order to maximize local memory cache hit.

**Information Distribution Service (IDS)** allows information exchange based on a publish/subscribe system. Some of the information disseminated through it includes: resource utilization, service/unit resolving updates, system status updates, log levels and log entries, global configuration, etc.

**Deployment Database Service (DDS)** maintains copies of VM images and service and Microservice binaries that are necessary to deploy software on the Pouches of the system.

**Log Gathering Service (LGS)** sorts and consolidates the logging information received from the Pouches, Services and Microservices.

**3.2 The IMS Telephony Application**

To study the advantages of a Hyper Heterogeneous Cloud-based approach (fully distributed and elastic deployment on heterogeneous platforms) versus a Node-based approach (functions constrained to dedicated hardware or virtual machine (VM)) in terms of telecommunication functionality, we built a simplified IMS on the Unity Cloud Microservices-based architecture with the goal of deploying it in a heterogeneous cloud infrastructure. This allowed us to select the distribution of the functions on the physical or virtual platform i.e. IMS functions can be fully distributed on a pool of compute resources (Cloud-based) or on a specific compute resource (Node-based) given a single software base, thus enabling a fair comparison. The simplified IMS functions are split amongst a number of communicating Microservices joined in a complete service chain (or call chain). Fig. 5 illustrates how the Microservices are linked in a complete service chain to provide a phone call between two subscribers.

The Microservices (also called Units) developed for the Unity Cloud IMS Telephony Application are listed below with notes as to which of the IMS functions they provide.

**SIP Handler (SIPh)** implements the SIP processing functions of the P-CSCF and the I-CSCF. It is the first Unit involved in a service setup scenario. It uses the Node Selector Service to figure out where the Call Session Unit should be instantiated and forwards it the received SIP messages.

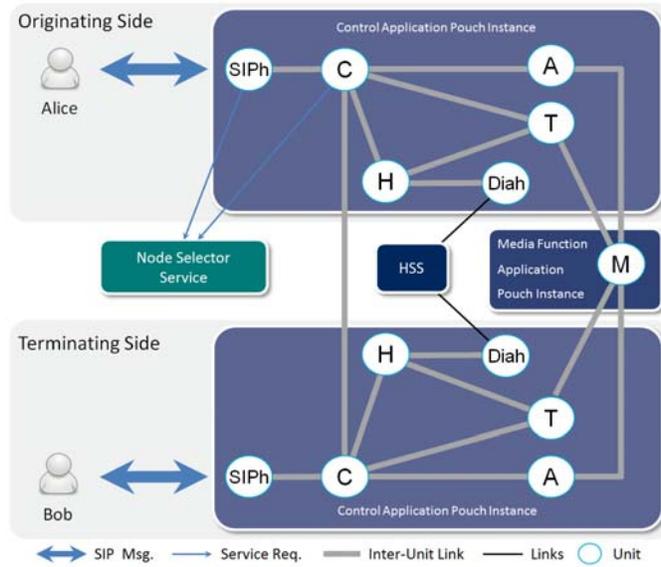

**Fig. 5.** Microservices involved in a typical two-party call scenario.

**Call Session (C)** performs the functionality of an S-CSCF. It handles the request coming from the UA, fetches the Subscriber profile based on service triggers and builds the appropriate service chain to provide the requested service. The C Unit is instantiated on request to handle the Subscriber service and is terminated when the service has completed e.g. during a call, it remains active until the SIP Bye message has been acknowledged. The C Unit makes use of the Node Selector Service in order to figure out where the terminating Call Session Unit should be instantiated.

**HSS Front-End (H)** is used to fetch a Subscriber profile. It is responsible for querying the HSS database in order to get this information.

**Diameter Handler (Diah)** is used by the H Unit as an interface that implements the diameter protocol towards the HSS in order to fetch a Subscriber profile.

**Anchor Point Controller (A)** covers the MRFC functionality controlling the Media Processor Unit as needed for the requested service and informs the interested Units of the availability of the functionality in the service chain. The A Unit's main function is to negotiate the media codec so that the UA can properly exchange media with the Media Processor Unit.

**Telephony Server (T)** provides telephony related features to the Subscriber. As an IMS MMTEL it can listen to DTMF activities to trigger supplementary services like ad-hoc conferences by adding another call leg to the current call. It is created by the C Unit on both the originating and terminating sides based on the Subscriber profile fetched via the H Unit; it connects to the M Unit to receive the media plane telephony events and to control the connectivity of the media plane.

**Media Processor (M)** is a dialog-based Microservice that handles the media plane of the call through RTP as an IMS MRFP would do. It provides point-to-point connectivity for basic 2-way calls and provides voice mixing in conference calls.

### 3.3 The Hardware Platform

We deployed our Microservices IMS Telephony application on two distinct platforms. The first deployment platform (Fig. 6.A) is based on a cluster made of eight Raspberry Pi's [16] (RPi). The benefits of this platform are twofold. Firstly, it is a cost effective way to have a 24/7 cloud we can experiment on and secondly, RPi being a simple single core computer, limits the number of variables required to consider while studying the system.

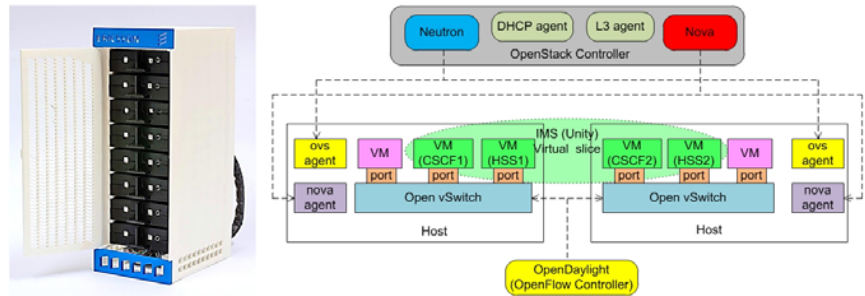

**Fig. 6.** (A) Eight Raspberry Pi boards Unity Cloud 3D printed Cabinet; and (B) Unity deployment on OpenStack.

The Unity Cloud RPi platform is built of:
- 8 Model B RPi stacked together in a custom made 3D printed cabinet where each RPi is set in a removable sliding tray.
- 8 custom-made RPi Daughter Boards enabling the display of information via 2 RGB LEDs and allowing input via a button.
- 1 Gigabit Ethernet switch providing the backbone network for the system.
- 1 Wi-Fi router providing access to UEs (hosting the UA) and providing the NAS functionality on a USB Storage Device.
- 1 Power Supply for the Cabinet.

The second deployment (Fig. 6.B) is on top of OpenStack deployed on an Ericsson Blade System (EBS) [17] consisting of 8 VMs (2 virtual cores and 2GB of RAM) deployed on 4 physical blades. An automatic orchestration mechanism is triggered to balance load of the blades though VM migrations.

## 4   Experimentation

The first experiment is carried out to demonstrate the compatibility and compliance between different cloud-based deployment platforms (RPi cluster and EBS VMs) regarding IMS telephony Microservice functions developed for the Unity Cloud.

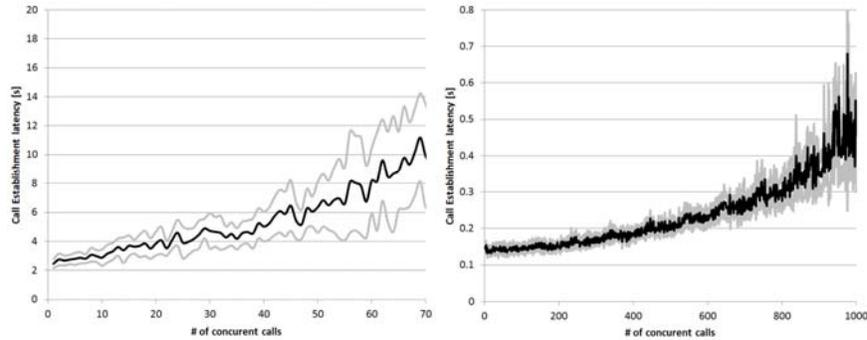

**Fig. 7.** Average and Standard Deviation of the Call Establishment Latency on RPi cluster (left) and on EBS VMs (right).

Our measurement consists of a collection of in-process logs which are collected in a file on the Unity Cloud. The open-source tool SIPp [18] has been used to generate SIP traffic.

We measure the delay from the reception of the SIP INVITE by the Unity Cloud until it is sent to the terminating UA (Fig. 7). This way we keep the measurement to the portion that is directly dependent on Unity Cloud processing and independent from UA delays. We also take measurements about the average CPU load on all Computing Units (CUs) against the number of concurrent calls being served by the system (Fig. 8). For all measurements we settled on:
- Call Rate: 30 2-way calls establishment / minute
- Call Duration: 3:20 minutes
- Subscribers: 200 registered users
- Background Registration: 20 re-registration / minute

As shown in Fig. 7, calls could be successfully established in both experimental platforms and the QoS characteristics show similar behavior. QoS characteristics are obviously better on the more powerful EBS but the QoS trend is similar to the RPi cluster. In Fig. 8, we notice similarities in the resource usage profile where average CPU usage increases relatively linearly with the number of concurrent calls.

In order to compare the proposed Microservice-based architecture (where the functions can be fully distributed on the available CUs) to the currently prominent Node-based architecture, (where the functions of a node are bound to a set of CUs) we conducted a set of experiments where the functions developed for the Unity Cloud were statically bound to a specific CU thus replicating the Node-based architecture (Table 1).

In a Node-based architecture the provisioning of the nodes needs to be perfectly engineered. However, since we had only a limited number of CUs available and didn't have proper methods to manually engineer the provisioning, we evaluated a number of configurations. These configurations and their functionality distribution are shown for the 8 available CUs in Table 1.

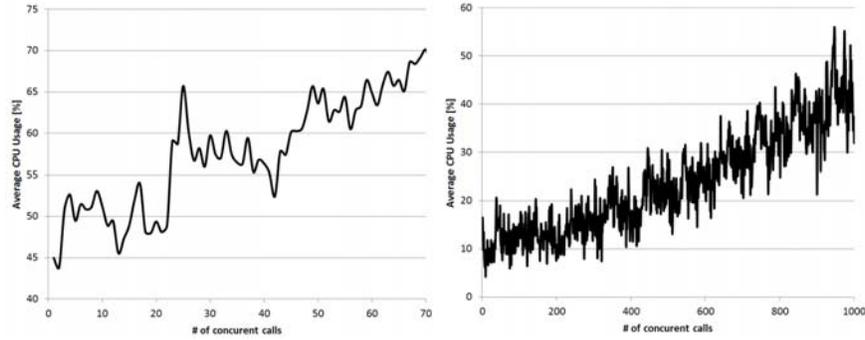

**Fig. 8.** Overall CPU usage Average for a specific load measured in the number of concurrent calls on the RPi cluster (left) and on EBS VMs (right).

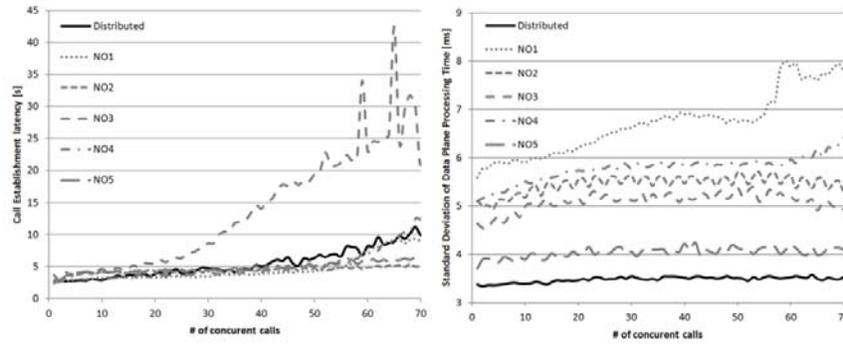

**Fig. 9.** Call Establishment Latency (left) and Data Plane std. dev. to the 20ms boundary processing time vs the Number Of Calls (right) for the different experimentation configurations.

**Table 1.** Experimental configurations for Node-based measurements: S (SIPh), N (NSS), H (H and Diah), all other Units as previously described.

| Config. | CU1 | CU2 | CU3 | CU4 | CU5 | CU6 | CU7 | CU8 |
|---------|-----|-----|-----|-----|-----|-----|-----|-----|
| NO1     | SN  | H   | C   | C   | A   | T   | M   | M   |
| NO2     | SNH | C   | C   | A   | T   | M   | M   | M   |
| NO3     | SN  | H   | C   | A   | T   | M   | M   | M   |
| NO4     | SN  | H   | C   | C   | T   | MA  | MA  | MA  |
| NO5     | SNH | C   | C   | T   | MA  | MA  | MA  | MA  |

As shown in Fig. 9 the distributed Cloud-based approach gives similar average control plane QoS characteristics compared to Node-based approach while avoiding the worst case scenario exposed by poorly engineered resource allocation. This is depicted in the Node-based deployment configuration NO3, where the C Unit processing is starved as it is deployed on a single CU. It is worth noting that the Cloud-based approach exhibits the best data plane QoS characteristics compared to all Node-based deployment configurations.

## 5   Discussion and Conclusion

Our deployment of a Microservices-based IMS telephony solution on different cloud platforms (RPi cluster and OpenStack/EBS platform) shows that this architecture enables one-time development of the business logic across multiple deployments on various platforms through modification of deployment configuration only. It demonstrates the possibility of defining an architecture supporting Cloud features, especially automatic scaling out of the business logic for the telecom sector. Fig. 7 and Fig. 8 show that the QoS of the application and platform (e.g., in terms of Call Establishment Delay and CPU consumption) are comparable in both RPi and EBS deployments. A larger number of calls can be handled on the EBS deployment but the trend of QoS characteristics stays the same. This suggests that we achieved our goal of defining a cloud-based software architecture that can be easily deployed on heterogeneous hardware clusters using a single application code base.

In Fig. 9 we observe that accurately allocating resources is required for each node in a Node-based system, and this must be done statically due to the static configuration of the Node-based system. For example, dramatic performance degradation is experienced on NO3 where the lack of resources allocated to the C Unit degrades the control plane latency in a very noticeable fashion. Aside from control plane Unit resource allocation, media plane resource allocation must also be considered. In contrast, the distributed Cloud-based approach allocates the same amount of resources to both media and control plane and also to each unit since distribution is based on CPU usage. As such, we observe better media plane performance and average control plane performance. Using our approach, the Microservices can be distributed and combined without location restrictions on VMs in order to efficiently use the available resources. This is an advantage compared to a Node-based deployment where functionality is bound to specific resources.

In conclusion, a distributed Cloud-based approach can provide automatic platform resource allocation which cannot be easily achieved by a Node-based architecture. Failure to properly engineer node resource allocation in a node-based architecture can lead to major impacts on performance. Node-based deployment also reduces the reliability of the overall system since if a node is deployed on only one CU and it fails, then the whole system fails and the service will remain unavailable until that function is restored. In the distributed cloud-based model, such a failure will terminate the services hosted on a single CU but the system will remain available to provide new service instances spread across other available CUs. The hyper-heterogeneity aspect of the proposed architecture enables us to deploy an application that is designed once across different cloud platforms, thus easing the job of telco vendors to deploy network functions on various operator-owned clouds. The hyper-heterogeneity aspect also allows tailoring of the deployment to take advantage of the benefits of specific platforms for a given application. For example, an accelerator-based cloud might be beneficial for media resource processing functions while a general-purpose cloud might be more appropriate for control information.

In the future it would be interesting to evaluate heterogeneous deployments where a system would be deployed on integrated clusters of different technologies e.g. bare metal server pool used for data plane processing and a VM-based cloud used for control plane processing.

**Acknowledgements.** This work is sponsored by Ericsson Canada Inc. where we would like to thank the team researchers developing the PoC: Marc-Olivier Arsenault, Gordon Bailey, Mario Bonja, Léo Collard, Alexis Grondin, Philippe Habib, Olivier Lemelin, Mahdy Nabaee, Maxime Nadeau, Fakher Oueslati and Joseph Siouffi.